# TelegramScrap: A comprehensive tool for scraping Telegram data

*Ergon Cugler de Moraes Silva*

Brazilian Institute of Information in
Science and Technology (IBICT)
Brasília, Federal District, Brazil

contato@ergoncugler.com
www.ergoncugler.com

## Abstract

[WhitePaper] The **TelegramScrap** tool provides a robust and versatile solution for extracting and analyzing data from Telegram channels and groups, addressing the increasing demand for efficient methods to study digital ecosystems. This white paper outlines the tool's development, capabilities, and applications in academic and scientific research, including studies on disinformation, political communication, and thematic patterns in online communities. Built with flexibility and user accessibility in mind, the tool allows researchers to customize scraping parameters, handle large datasets, and produce structured outputs in formats such as Excel and Parquet. Its modular architecture, real-time progress tracking, and error-handling mechanisms ensure reliability and scalability for diverse research needs. Emphasizing ethical data collection, the tool aligns with Telegram's terms of service and data privacy regulations, encouraging responsible use. Released under an open-source license, **TelegramScrap** invites the academic community to explore, adapt, and improve the tool while providing appropriate credit. This paper demonstrates the tool's impact through its application in multiple studies, showcasing its potential to advance computational social science and enhance understanding of digital interactions and societal trends.

**[ Code available on GitHub: https://github.com/ergoncugler/web-scraping-telegram ]**

## 1. Introduction

Telegram has positioned itself as one of the most versatile communication platforms of the digital age, renowned for its robust privacy features, scalability, and diverse functionalities, such as encrypted messaging, public channels, and bot integrations. Its decentralized structure has made it a favored platform for fostering communities and enabling large-scale information dissemination, particularly in regions where access to traditional media is restricted. Rudnik and Rönnblom (2024) explored its role in political mobilization in Belarus and Russia, emphasizing its potential for fostering civic participation and alternative discourse in constrained environments. Similarly, Lawrence (2024) highlighted how Telegram and similar platforms are reshaping political communication, becoming critical tools for grassroots activism and public engagement in Western democracies.

Also, the educational and collaborative capabilities of Telegram have gained prominence, with researchers exploring its use in various academic and professional contexts. Osondu, Anugwo, and Ugama (2024) demonstrated its effectiveness in fostering collaborative learning and improving academic outcomes among tertiary-level students. Efremov and Lukinova (2024), as an example, examined its role in shaping digital communication etiquette, underscoring its relevance for both personal and professional interactions. Moreover, Aduma and Ntaka (2024) highlighted the platform's utility in entrepreneurial education, offering a bridge for digital innovation and capacity building in under-resourced areas. Studies by Bredikhin and Avdeev (2024) also examined Telegram's contribution to understanding communication dynamics in complex socio-political settings, particularly in regions experiencing heightened conflict or instability.

Beyond communication and education, Telegram has emerged as an essential resource for researchers seeking to understand digital interactions and societal trends. The development of tools to collect and analyze data from Telegram channels and groups has enabled groundbreaking studies in computational social science. Illia and Tetiana (2024) focused on software designed for processing text data from Telegram, highlighting its utility for analyzing user-generated content on a large scale. Ebrahimi and Soltanifar (2024) explored the platform's role in disseminating fake news, shedding light on the mechanisms through which misinformation spreads in digital ecosystems. Furthermore, Admassie and Melesse (2024) investigated Telegram's potential in facilitating collaborative medical research, particularly in remote regions, underscoring its adaptability to a wide array of fields.

Whether in political mobilization, education, or digital behavior analysis, Telegram serves as a key enabler of modern scholarship and professional engagement. The platform's technical capabilities and the analytical tools developed around it will continue to expand the boundaries of research, offering new insights into complex social and digital phenomena.

Building upon the diverse applications highlighted in the academic literature, this white paper showcases the practical impact of the Telegram scraping code developed for research purposes. By enabling data collection and analysis from Telegram groups and channels, the tool has been instrumental in advancing empirical studies across various domains. Notably, it has supported investigations into disinformation dynamics, political discourses, and thematic convergence in conspiracy theory communities. Below, we outline key examples where this tool has contributed to groundbreaking research, demonstrating its versatility and relevance in addressing pressing societal and scholarly questions.

Just to list some examples of code usage, it was used in the paper "Informational Co-option against Democracy: Comparing Bolsonaro's Discourses about Voting Machines with the Public Debate" (Silva & Oliveira, 2023). It was also used in "Institutional Denialism From the President's Speeches to the Formation of the Early Treatment Agenda (Off Label) in the COVID-19 Pandemic in Brazil" (Silva, 2023b). Moreover, the code facilitated research in "Catalytic Conspiracism: Exploring Persistent Homologies Time Series in the Dissemination of Disinformation in Conspiracy Theory Communities on Telegram" (Rocha, Silva & Mielli, 2024) and "Conspiratorial Convergence: Comparing Thematic Agendas Among Conspiracy Theory Communities on Telegram Using Topic Modeling" (Silva & Máximo, 2024). Also, it was pivotal in the study "Informational Disorder and Institutions Under Attack: How Did Former President Bolsonaro's Narratives Against the Brazilian Judiciary Between 2019 and 2022 Manifest?" (Silva, Castro, Santos, 2023).

Furthermore, the code was utilized in several technical notes, as we can see in "Technical Note #16 – Disinformation about Electronic Voting Machines Persists Outside Election Periods" (Silva, 2023c). It was also employed in "Technical Note #18 – Electoral Fraud Discourse in Argentina on Telegram and Twitter" (Silva & Ortellado, 2023). The code also contributed to the analysis in the technical note "Bashing and Praising Public Servants and Bureaucrats During the Bolsonaro Government (2019 - 2022)" (Silva, Lotta, Fernandez, Seidi & Oliveira, 2024). Additionally, it was used in "Technical Note 2: The Digital Territory of Milei's Followers: From Commerce to Politics" (Vargas, Hernández, Silva, Foltran, 2023).

This tool was also used in a series of 7 articles published on ArXiv about conspiracy theory communities, consolidating a compilation of analysis on the topic in Brazil: 1.) "Antivax and off-label medication communities on Brazilian Telegram: between esotericism as a gateway and the monetization of false miraculous cures" (Silva, 2024a); 2.) "Climate change denial and anti-science communities on Brazilian Telegram: climate disinformation as a gateway to broader conspiracy networks" (Silva, 2024d); 3.) "Anti-woke agenda, gender issues, revisionism and hate speech communities on Brazilian Telegram: from harmful reactionary speech to the crime of glorifying Nazism and Hitler" (Silva, 2024c); 4.)

"Apocalypse, survivalism, occultism and esotericism communities on Brazilian Telegram: when faith is used to sell quantum courses and open doors to harmful conspiracy theories" (Silva, 2024b); 5.) "UFO, universe, reptilians and creatures communities on Brazilian Telegram: when the sky is not the limit and conspiracy theories seek answers beyond humanity" (Silva, 2024h); 6.) "Flat-earth communities on Brazilian Telegram: when faith is used to question the existence of gravity as a physics phenomenon" (Silva, 2024e); and 7.) "New world order, globalism and QAnon communities on Brazilian Telegram: how conspiracism opens doors to more harmful groups" (Silva, 2024g).

Moreover, this tool has been applied in procedural decision-making contexts, such as during the Joint Parliamentary Commission of Inquiry on the events of January 8, 2023. Data gathered through this tool was instrumental in substantiating the subpoenas of individuals involved in the criminal invasion of Brasília, Brazil, on that date (Senado Federal, 2023; Psol na Câmara, 2023). The capacity to systematically collect and analyze large-scale data enabled authorities to identify and target specific actors within the investigative framework, showcasing the tool's utility in legal and political proceedings. In addition, this technology has been utilized for tracking criminals and addressing cybercrimes in Italy. Researchers affiliated with the DarkLab of Ethical Hacking highlighted its role in uncovering and prosecuting offenders involved in malicious online activities, particularly through the analysis of data leaks and cyber threats (RedHot Cyber, 2024). These examples underscore the versatility of the tool, proving its effectiveness in diverse applications ranging from public security to cybercrime investigation, while advancing digital forensics and procedural justice.

## 2. Step-by-step usage

Just some introductory tips:

- ➔ **Google Colab runtime limit:** Google Colab typically crashes after running this code for around 6 hours and 20 minutes (which is 22,800 seconds). Therefore, set a limit within this timeframe to avoid interruptions.
- ➔ **Telegram API soft ban:** The Telegram API usually imposes a 24-hour soft ban after scraping more than 200 channels or groups. However, there seems to be no limit on the number of messages scraped from fewer communities. To avoid the ban, scrape large amounts of content from blocks of up to 150-200 communities at a time, even if you extract entire months of data from each one or just days.

- **Using Google Colab for async operations:** One advantage of using Google Colab is the ability to run async without needing to define them within an async def. If you plan to use PyCharm or another IDE, consider adapting the code with an async def.
- **Handling JSON in 'Comments List' column:** The 'Comments List' column stores comments in a JSON list format. Remember to decode this JSON when converting to a spreadsheet or presenting the data. It is also important to consider that if the list of comments is too long, it will not fit in the Excel cell and it is for this reason (among other features) that we provide the parquet file with JSON, to avoid content breaks.

### I. [ Required ] Set up your credentials once

The script is designed to facilitate data scraping from Telegram channels or groups using a structured step-by-step process. In the first step, users need to set up their credentials, which is a one-time configuration. This involves entering the required information: username, phone, api_id, and api_hash. The username should be provided without the "@" symbol, and the phone number must be formatted internationally, for example, "+5511999999999". The api_id and api_hash are essential for authentication and can only be generated through Telegram's app creation page, accessible at https://my.telegram.org/apps. After filling in these fields, the user simply runs the cell to initialize the credentials. These settings do not need to be updated unless the user decides to change their Telegram account.

**Figure 1** - First cell to be evaluated

**Source:** Own elaboration (2024).

### II. [ Required ] Adjust every time you want to use it

In the second step, users configure the parameters for each scraping session. This section allows users to define the specific conditions for data collection. The first parameter is the list of Telegram channels or groups to scrape. These should be written as @ChannelName or the full URL, such as https://t.me/ChannelName, separated by commas. Links from the Telegram web version, such as "https://web.telegram.org", should not be used. The second parameter is the date range, defined by date_min and date_max, which specifies the period from which messages will be collected. The dates should be formatted as YYYY-MM-DD. Users also need to name the output file for the collected data, providing clarity for future use. An optional keyword search field allows users to filter messages by specific terms; if left blank, all messages within the specified range will be scraped. Additionally, users can set a maximum number of messages to scrape, ensuring the process remains manageable. The timeout duration, which is the maximum time the script will run, should also be defined. This value should not exceed six hours (21,600 seconds), as Google Colab sessions automatically terminate after that time. Lastly, users choose the format of the output file, either excel for accessibility or parquet for advanced data processing.

**Figure 2** - Second cell with search queries

**Source:** Own elaboration (2024).

### III. [ Required ] Start Telegram scraping

The third step involves running the scraping process. During this step, the script initiates data collection based on the previously defined parameters. Users should be aware that Telegram may request a verification code during the process, which will need to be entered promptly. The script collects messages along with their metadata, such as views, shares, and reactions, as well as comments when available. Progress updates are displayed throughout the session, showing the percentage completed, elapsed time, and estimated time remaining. To minimize the risk of data loss in extended sessions, backup files are created automatically after every 1,000 messages processed. These backup files are saved in the specified output format, ensuring users have access to intermediate results even if the session is interrupted. Once the scraping process concludes, the script compiles a final file containing all collected data, ready for download.

**Figure 3 -** Third cell where Metadata is extracted

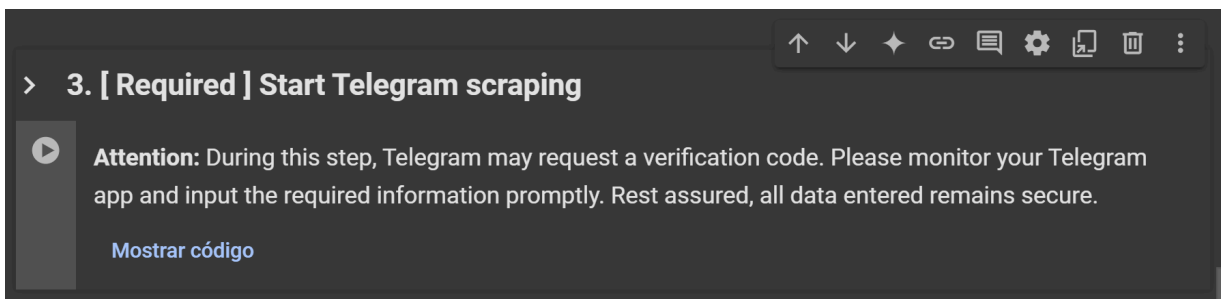

**Source:** Own elaboration (2024).

## 3. Code

### I. Set up your credentials once

**Table 01. First cell approaches and code**

| Approach description | Code description |
|---|---|
| It utilizes the Telethon library, a robust tool for programmatically interfacing with Telegram. Users are prompted to input their Telegram username, phone number, api_id, and api_hash, which are essential credentials generated from Telegram's app creation page (https://my.telegram.org/apps). These details establish the connection to the user's Telegram account. In addition to Telethon, the script imports foundational libraries like datetime for managing timestamps, pandas for data manipulation, json and re for processing and cleaning textual data, and time for tracking execution durations. Google Colab's files module is included to facilitate file downloads directly to the user's machine. This cell is run only once unless credentials need updating. | `# Install the Telethon library for Telegram API interactions`<br>`!pip install -q telethon`<br><br>`# Initial imports`<br>`from datetime import datetime, timezone`<br>`import pandas as pd`<br>`import time`<br>`import json`<br>`import re`<br><br>`# Telegram imports`<br>`from telethon.sync import TelegramClient`<br><br>`# Google Colab imports`<br>`from google.colab import files`<br><br>`# Setup / change only the first time you use it`<br>`# @markdown **1.1.** Your Telegram account username (just 'abc123', not '@'):`<br>`username = 'your_username' # @param {type:"string"}`<br>`# @markdown **1.2.** Your Telegram account phone number (ex: '+5511999999999'):`<br>`phone = '+5511999999999' # @param {type:"string"}`<br>`# @markdown **1.3.** Your API ID, it can be only` |

| Approach description | Code description |
|---|---|
| | ```
generated from https://my.telegram.org/apps:
api_id = '11111111' # @param {type:"string"}
# @markdown **1.4.** Your API hash, also from
https://my.telegram.org/apps:
api_hash = '1a1a1a1a1a1a1a1a1a1a1a1a1a1a1a1a' #
@param {type:"string"}
``` |

**Source:** Own elaboration (2024).

## II. *Adjust every time you want to use it*

### Table 02. Second cell approaches and code

| Approach description | Code description |
|---|---|
| It provides a flexible interface for defining the parameters of each Telegram data scraping session, offering full customization to meet diverse research needs. This cell leverages essential libraries like datetime to manage date ranges and pandas for structured data manipulation, ensuring precise processing of user-defined configurations. Users are prompted to input a list of Telegram channels or groups in the format @ChannelName or full URLs such as https://t.me/ChannelName. Dynamic string manipulation techniques (.split, .strip) handle the separation and cleaning of input values, while links from incompatible platforms like https://web.telegram.org are excluded to ensure seamless interaction with Telegram's API. This cell allows users to specify a date range (date_min and date_max) to filter messages from specific periods. Dates are automatically converted to UTC using datetime.fromisoformat, making this feature particularly valuable for time-sensitive analyses, such as examining discussions surrounding political events or public policies. Users also assign a name to the output file (file_name) to organize multiple datasets effectively. An optional keyword filter can be applied to focus on messages containing specific terms, while leaving the field blank enables the extraction of all available messages. To ensure the process remains manageable, users can define a | ```
# Setup / change every time to define scraping
parameters

# @markdown **2.1.** Here you put the name of the
channel or group that you want to scrape, as an
example, play: '@LulanoTelegram' or
'https://t.me/LulanoTelegram'. Do not use:
'https://web.telegram.org/a/#-1001249230829' or
'-1001249230829'. **Just write the `channel names`
always separated by commas (,):**
channels = "@LulanoTelegram, @jairbolsonarobrasil,
@Other_Channel_Name" # @param {type:"string"}
channels = [channel.strip() for channel in
channels.split(",")]

# @markdown **2.2.** Here you can select the `time
window` you would like to extract data from the
listed communities:
date_min = '2024-10-15' # @param {type:"date"}
date_max = '2025-01-15' # @param {type:"date"}

date_min =
datetime.fromisoformat(date_min).replace(tzinfo=tim
ezone.utc)
date_max =
datetime.fromisoformat(date_max).replace(tzinfo=tim
ezone.utc)

# @markdown **2.3.** Choose a `name` for the final
file you want to download as output:
file_name = 'Test' # @param {type:"string"}

# @markdown **2.4.** `Keyword` to search, **leave
empty if you want to extract all messages from the
channel(s):**
key_search = '' # @param {type:"string"}

# @markdown **2.5.** **Maximum** `number of
messages` to scrape (only use if you want a
specific limit, otherwise leave a high number to
scrape everything):
max_t_index = 1000000   # @param {type:"integer"}

# @markdown **2.6.** `Timeout in seconds` (never
leave it longer than 6 hours, that is 21600
seconds, as Google Colab deactivates itself after
``` |

| Approach description | Code description |
|---|---|
| maximum number of messages to scrape (max_t_index) and set a timeout limit in seconds. This timeout ensures the session concludes within Google Colab's six-hour runtime restriction (21,600 seconds). The cell also provides the option to choose between two output file formats: Excel for accessibility and ease of use, or Parquet, which is optimized for handling large datasets and is widely adopted in big data workflows. | ```that time):
time_limit = 21600 # @param {type:"integer"}

# @markdown **2.7.** Choose the format of the final file you want to download. If you are a first-time user, choose `Excel`. If you have advanced skills, you can use `Parquet`:
File = 'excel' # @param ["excel", "parquet"]``` |

**Source:** Own elaboration (2024).

### III. Start Telegram scraping

**Table 03. Third cell approaches and code**

| Approach description | Code description |
|---|---|
| It executes the core data collection process by integrating the configurations and libraries defined in the previous steps. It employs the TelegramClient from the Telethon library to interact programmatically with Telegram's API. The scraping process is asynchronous, enabling efficient performance even when handling large datasets on resource-limited platforms like Google Colab. The cell initializes global variables such as a list for storing collected data (data), a message counter (t_index), and a timestamp to track the session's start time (start_time). These foundational elements facilitate progress tracking and ensure organized data handling throughout the session. One of the cell's key functionalities is its ability to process messages and their metadata, including views, reactions, shares, and any associated media. Using an asynchronous loop, the script iterates through messages within the specified date range, also capturing and processing comments nested under individual messages. It employs custom functions like remove_unsupported_characters to clean text by removing invalid XML | ```python
data = []  # List to store scraped data
t_index = 0   # Tracker for the number of messages processed
start_time = time.time()  # Record the start time for the scraping session

# Function to remove invalid XML characters from text
def remove_unsupported_characters(text):
    valid_xml_chars = (
        "[^\u0009\u000A\u000D\u0020-\uD7FF\uE000-\uFFFD"
        "\U00010000-\U0010FFFF]"
    )
    cleaned_text = re.sub(valid_xml_chars, '', text)
    return cleaned_text

# Function to format time in days, hours, minutes, and seconds
def format_time(seconds):
    days = seconds // 86400
    hours = (seconds % 86400) // 3600
    minutes = (seconds % 3600) // 60
    seconds = seconds % 60
    return f'{int(days):02}:{int(hours):02}:{int(minutes):02}:{int(seconds):02}'

# Function to print progress of the scraping process
def print_progress(t_index, message_id, start_time, max_t_index):
    elapsed_time = time.time() - start_time
    current_progress = t_index / (t_index + message_id) if (t_index + message_id) <=
``` |



| Approach description | Code description |
|---|---|
| characters, ensuring the final dataset is free of encoding errors. To enhance user experience, progress is displayed in real-time through functions like format_time and print_progress, which provide updates on elapsed time, completion percentage, and estimated remaining time. The cell incorporates robust data security measures by automatically generating backup files every 1,000 messages. These backups are dynamically named to include the number of collected messages and the channel identifier, ensuring traceability. Depending on the user's preference, the files are saved in either Excel or Parquet format, providing flexibility for both casual and advanced users. If errors occur during message or comment processing, exceptions are handled gracefully, allowing the script to continue execution without significant interruptions. Time management is another critical feature of this cell. The script monitors the session's runtime, terminating automatically if the predefined timeout limit is reached, thus adhering to Google Colab's runtime restrictions. Once the scraping session concludes, the script compiles all collected data into a final output file and makes it available for direct download. This is facilitated by Google Colab's files module, streamlining the transfer of results to the user's local device. The cell's architecture demonstrates a sophisticated integration of asynchronous programming, real-time progress tracking, and file management, ensuring efficient operation even in complex scenarios. Its modular design supports easy adaptation for future studies, such as social network analyses, monitoring misinformation, or investigating community dynamics. | ```python
max_t_index else t_index / max_t_index
    percentage = current_progress * 100
    estimated_total_time = elapsed_time / current_progress
    remaining_time = estimated_total_time - elapsed_time

    elapsed_time_str = format_time(elapsed_time)
    remaining_time_str = format_time(remaining_time)

    print(f'Progress: {percentage:.2f}% | Elapsed Time: {elapsed_time_str} | Remaining Time: {remaining_time_str}')

# Normalize File variable to avoid issues
File = re.sub(r'[^a-z]', '', File.lower())  # Converts to lowercase and removes non-alphabetic characters

# Scraping process
for channel in channels:
    if t_index >= max_t_index:
        break

    if time.time() - start_time > time_limit:
        break

    loop_start_time = time.time()

    try:
        c_index = 0
        async with TelegramClient(username, api_id, api_hash) as client:
            async for message in client.iter_messages(channel, search=key_search):
                try:
                    if date_min <= message.date <= date_max:

                        # Process comments of the message
                        comments_list = []
                        try:
                            async for comment_message in client.iter_messages(channel, reply_to=message.id):
                                comment_text = comment_message.text.replace("'", '"')

                                comment_media = 'True' if comment_message.media else 'False'

                                comment_emoji_string = ''
                                if comment_message.reactions:
                                    for reaction_count in comment_message.reactions.results:
                                        emoji = reaction_count.reaction.emoticon
                                        count = str(reaction_count.count)
``` |

| Approach description | Code description |
|---|---|
|  | ```python
                                        comment_emoji_string += emoji + " " + count + " "
                                        comment_date_time = comment_message.date.strftime('%Y-%m-%d %H:%M:%S')

                                        comments_list.append({
                                            'Type': 'comment',
                                            'Comment Group': channel,
                                            'Comment Author ID': comment_message.sender_id,
                                            'Comment Content': comment_text,
                                            'Comment Date': comment_date_time,
                                            'Comment Message ID': comment_message.id,
                                            'Comment Author': comment_message.post_author,
                                            'Comment Views': comment_message.views,
                                            'Comment Reactions': comment_emoji_string,
                                            'Comment Shares': comment_message.forwards,
                                            'Comment Media': comment_media,
                                            'Comment Url': f'https://t.me/{channel}/{message.id}?comment={comment_message.id}'.replace('@', ''),
                                        })
                        except Exception as e:
                            comments_list = []
                            print(f'Error processing comments: {e}')

                        # Process the main message
                        media = 'True' if message.media else 'False'

                        emoji_string = ''
                        if message.reactions:
                            for reaction_count in message.reactions.results:
                                emoji = reaction_count.reaction.emoticon
                                count = str(reaction_count.count)
                                emoji_string += emoji + " " + count + " "

                        date_time = message.date.strftime('%Y-%m-%d %H:%M:%S')
                        cleaned_content = remove_unsupported_characters(message.text)
                        cleaned_comments_list = remove_unsupported_characters(json.dumps(comments_list))

                        data.append({
                            'Type': 'text',
``` |



| Approach description | Code description |
|---|---|
|  | ```python
                                'Group': channel,
                                'Author ID': message.sender_id,
                                'Content': cleaned_content,
                                'Date': date_time,
                                'Message ID': message.id,
                                'Author': message.post_author,
                                'Views': message.views,
                                'Reactions': emoji_string,
                                'Shares': message.forwards,
                                'Media': media,
                                'Url': f'https://t.me/{channel}/{message.id}'.replace('@', ''),
                                'Comments List': cleaned_comments_list,
                            })

                            c_index += 1
                            t_index += 1

                            # Print progress
                            print(f'{"-" * 80}')
                            print_progress(t_index, message.id, start_time, max_t_index)
                            current_max_id = min(c_index + message.id, max_t_index)
                            print(f'From {channel}: {c_index:05} contents of {current_max_id:05}')
                            print(f'Id: {message.id:05} / Date: {date_time}')
                            print(f'Total: {t_index:05} contents until now')
                            print(f'{"-" * 80}\n\n')

                            if t_index % 1000 == 0:
                                if File == 'parquet':
                                    backup_filename = f'backup_{file_name}_until_{t_index:05}_{channel}_ID{message.id:07}.parquet'
                                    pd.DataFrame(data).to_parquet(backup_filename, index=False)
                                elif File == 'excel':
                                    backup_filename = f'backup_{file_name}_until_{t_index:05}_{channel}_ID{message.id:07}.xlsx'
                                    pd.DataFrame(data).to_excel(backup_filename, index=False, engine='openpyxl')

                            if t_index >= max_t_index:
                                break

                            if time.time() - start_time > time_limit:
                                break

                        elif message.date < date_min:
``` |

| Approach description | Code description |
|---|---|
| | ```
                    break

            except Exception as e:
                print(f'Error processing message: {e}')

        print(f'\n\n##### {channel} was ok with {c_index:05} posts #####\n\n')

        df = pd.DataFrame(data)
        if File == 'parquet':
            partial_filename = f'complete_{channel}_in_{file_name}_until_{t_index:05}.parquet'
            df.to_parquet(partial_filename, index=False)
        elif File == 'excel':
            partial_filename = f'complete_{channel}_in_{file_name}_until_{t_index:05}.xlsx'
            df.to_excel(partial_filename, index=False, engine='openpyxl')
        # files.download(partial_filename)

    except Exception as e:
        print(f'{channel} error: {e}')

    loop_end_time = time.time()
    loop_duration = loop_end_time - loop_start_time

    if loop_duration < 60:
        time.sleep(60 - loop_duration)

print(f'\n{"-" * 50}\n#Concluded! #{t_index:05} posts were scraped!\n{"-" * 50}\n\n\n\n')
df = pd.DataFrame(data)
if File == 'parquet':
    final_filename = f'FINAL_{file_name}_with_{t_index:05}.parquet'
    df.to_parquet(final_filename, index=False)
elif File == 'excel':
    final_filename = f'FINAL_{file_name}_with_{t_index:05}.xlsx'
    df.to_excel(final_filename, index=False, engine='openpyxl')
files.download(final_filename)
``` |

**Source:** Own elaboration (2024).

## 4. Conclusions

The **TelegramScrap** tool proposes a comprehensive solution for extracting, organizing, and analyzing data from Telegram channels and groups, addressing the growing demand for effective tools to navigate digital platforms. By integrating user-friendly functionalities, the tool bridges the gap between raw data availability and actionable insights.



It empowers researchers, journalists, and analysts to explore nuanced digital behaviors, monitor disinformation campaigns, and investigate thematic patterns in online communities.

One of its core strengths lies in its adaptability, catering to diverse research needs and methodologies. The tool's modular architecture allows for seamless customization, enabling users to tailor scraping parameters, output formats, and analytical focuses according to their specific objectives. Its real-time progress tracking, and automated backups enhance reliability, making it a valuable asset for longitudinal studies and high-volume data extraction.

Furthermore, the tool embodies an ethical approach to data collection, emphasizing transparency, user accountability, and adherence to platform regulations. It encourages responsible use by highlighting limitations such as API restrictions, runtime considerations, and the importance of respecting user privacy. These safeguards ensure that the tool is not only effective but also aligned with ethical standards and legal compliance.

Its use is highly encouraged and recommended for academic and scientific research, content analysis, sentiment analysis, and speech analysis. While it is free to use and modify, the responsibility for its use and any modifications lies with the user. Feel free to explore, utilize, and adapt the code to suit your needs, but please ensure you comply with Telegram's terms of service and data privacy regulations. This tool is released under a free and open-source license. When using or modifying the tool, please ensure to provide appropriate credit and citation. Referencing the tool in your research is appreciated and contributes to its continued development and improvement.

## 5. References

### 5.1. Tool

[ english version / português abaixo ]

### 5.3. *Some papers that used the code*

[ english version / português abaixo ]Rocha, I., Silva, E. C. M., & Mielli, R. V. (2024). **Catalytic conspiracism:** Exploring persistent homologies time series in the dissemination of disinformation in conspiracy theory communities on Telegram. Trabalho apresentado no 14º Encontro da Associação Brasileira de Ciência Política (ABCP), Salvador, Brasil. Available at: https://www.abcp2024.sinteseeventos.com.br/trabalho/view?ID_TRABALHO=687.

Senado Federal. (2023). **Relatório da Comissão Parlamentar Mista de Inquérito sobre os atos de 8 de janeiro de 2023.** Available at: https://legis.senado.leg.br/atividade/comissoes/comissao/2606/mna/relatorios.

Silva, E. C. M. (2024a). **Antivax and off-label medication communities on Brazilian Telegram:** Between esotericism as a gateway and the monetization of false miraculous cures. arXiv preprint. Available at: https://doi.org/10.48550/arXiv.2408.15308.

Silva, E. C. M. (2024b). **Apocalypse, survivalism, occultism and esotericism communities on Brazilian Telegram:** When faith is used to sell quantum courses and open doors to harmful conspiracy theories. arXiv preprint. Available at: https://doi.org/10.48550/arXiv.2409.03130.

Silva, E. C. M. (2024c). **Anti-woke agenda, gender issues, revisionism and hate speech communities on Brazilian Telegram:** From harmful reactionary speech to the crime of glorifying Nazism and Hitler. arXiv preprint. Available at: https://doi.org/10.48550/arXiv.2409.00325.

Silva, E. C. M. (2024d). **Climate change denial and anti-science communities on Brazilian Telegram:** Climate disinformation as a gateway to broader conspiracy networks. arXiv preprint. Available at: https://doi.org/10.48550/arXiv.2408.15311.

Silva, E. C. M. (2024e). **Flat-earth communities on Brazilian Telegram:** When faith is used to question the existence of gravity as a physics phenomenon. arXiv preprint. Available at: https://doi.org/10.48550/arXiv.2409.03800.

Silva, E. C. M. (2023b). **Institutional denialism:** From the president's speeches to the formation of the early treatment agenda (off label) in the COVID-19 pandemic in Brazil. Anais do Encontro Nacional de Ensino e Pesquisa do Campo de Públicas, 5. Available at: https://anepecp.org/ojs/index.php/br/article/view/561.

Silva, E. C. M. (2024g). **New world order, globalism and QAnon communities on Brazilian Telegram:** How conspiracism opens doors to more harmful groups. arXiv preprint. Available at: https://doi.org/10.48550/arXiv.2409.12983.

Silva, E. C. M. (2023c, Mar 22). *Nota Técnica #16 – Desinformação sobre urnas eletrônicas persiste fora dos períodos eleitorais.* Monitor do Debate Político no Meio Digital. Available at: https://www.monitordigital.org/2023/05/18/nota-tecnica-16-desinformacao-sobre-urnas-eletronicas-persiste-fora-dos-periodos-eleitorais/.

Silva, E. C. M. (2024h). **UFO, universe, reptilians and creatures communities on Brazilian Telegram:** When the sky is not the limit and conspiracy theories seek answers beyond humanity. arXiv preprint. Available at: https://doi.org/10.48550/arXiv.2409.02117.

## 6. Author biography

**Ergon Cugler de Moraes Silva** has a Master's degree in Public Administration and Government (FGV), a Postgraduate MBA in Data Science & Analytics (USP), a Bachelor's degree in Public Policy Management (USP), and is currently pursuing a Postgraduate degree in Data Science for Social and Business Analytics at the University of Barcelona. He is associated with More in Common (Brazil), collaborates with the Interdisciplinary Observatory of Public Policies (OIPP USP), the Study Group on Technology and Innovations in Public Management (GETIP USP), the Monitor of Political Debate in the Digital Environment (Monitor USP), and the Working Group on Strategy, Data and Sovereignty of the Study and Research Group on International Security of the Institute of International Relations of the University of Brasília (GEPSI UnB). He is also a researcher at the Brazilian Institute of



Information in Science and Technology (IBICT), where he works for the Federal Government on strategies against disinformation. São Paulo, São Paulo, Brazil. Website: https://ergoncugler.com/.



# TelegramScrap: Uma ferramenta abrangente para extração de dados do Telegram


*Ergon Cugler de Moraes Silva*

Instituto Brasileiro de Informação
em Ciência e Tecnologia (IBICT)
Brasília, Distrito Federal, Brasil

contato@ergoncugler.com
www.ergoncugler.com



**Resumo**

[WhitePaper] A ferramenta **TelegramScrap** oferece uma solução robusta e versátil para extração e análise de dados de canais e grupos do Telegram, atendendo à crescente demanda por métodos eficientes para o estudo de ecossistemas digitais. Este white paper detalha o desenvolvimento, as capacidades e as aplicações da ferramenta em pesquisas acadêmicas e científicas, incluindo estudos sobre desinformação, comunicação política e padrões temáticos em comunidades online. Desenvolvida com foco na flexibilidade e acessibilidade para o usuário, a ferramenta permite que pesquisadores personalizem parâmetros de extração, gerenciem grandes volumes de dados e produzam saídas estruturadas em formatos como Excel e Parquet. Sua arquitetura modular, o acompanhamento em tempo real do progresso e os mecanismos de tratamento de erros garantem confiabilidade e escalabilidade para atender às diversas necessidades de pesquisa. Com ênfase na coleta ética de dados, a ferramenta está alinhada aos termos de serviço do Telegram e às regulamentações de privacidade de dados, promovendo o uso responsável. Lançada sob uma licença de código aberto, o **TelegramScrap** convida a comunidade acadêmica a explorar, adaptar e aprimorar a ferramenta, assegurando o devido crédito ao utilizá-la. Este paper demonstra o impacto da ferramenta por meio de sua aplicação em múltiplos estudos, destacando seu potencial para impulsionar as ciências sociais computacionais e aprofundar a compreensão das interações digitais e das tendências sociais.

**[ Código disponível no GitHub: https://github.com/ergoncugler/web-scraping-telegram ]**


## 1. Introdução

O Telegram posicionou-se como uma das plataformas de comunicação mais versáteis da era digital, conhecido por suas robustas funcionalidades de privacidade, escalabilidade e diversas opções, como mensagens criptografadas, canais públicos e integrações com bots. Sua estrutura descentralizada tornou-o uma plataforma preferida para fomentar comunidades e possibilitar a disseminação de informações em larga escala, especialmente em regiões onde o acesso à mídia tradicional é restrito. Rudnik e Rönnblom (2024) exploraram seu papel na



mobilização política na Bielorrússia e na Rússia, enfatizando seu potencial para promover participação cívica e discursos alternativos em ambientes restritos. Da mesma forma, Lawrence (2024) destacou como o Telegram e plataformas semelhantes estão transformando a comunicação política, tornando-se ferramentas cruciais para o ativismo de base e o engajamento público nas democracias ocidentais.

Além disso, as capacidades educacionais e colaborativas do Telegram têm ganhado destaque, com pesquisadores explorando seu uso em diversos contextos acadêmicos e profissionais. Osondu, Anugwo e Ugama (2024) demonstraram sua eficácia em promover aprendizado colaborativo e melhorar os resultados acadêmicos entre estudantes de nível superior. Efremov e Lukinova (2024), por exemplo, examinaram seu papel na formação de uma etiqueta de comunicação digital, ressaltando sua relevância tanto para interações pessoais quanto profissionais. Ademais, Aduma e Ntaka (2024) destacaram a utilidade da plataforma na educação empreendedora, oferecendo uma ponte para a inovação digital e o desenvolvimento de capacidades em áreas com recursos limitados. Estudos de Bredikhin e Avdeev (2024) também analisaram a contribuição do Telegram para compreender dinâmicas de comunicação em contextos sociopolíticos complexos, particularmente em regiões que enfrentam conflitos ou instabilidade.

Além da comunicação e da educação, o Telegram emergiu como um recurso essencial para pesquisadores que buscam entender interações digitais e tendências sociais. O desenvolvimento de ferramentas para coletar e analisar dados de canais e grupos do Telegram permitiu estudos inovadores em ciências sociais computacionais. Illia e Tetiana (2024) focaram em softwares projetados para processar dados textuais do Telegram, destacando sua utilidade para análise de conteúdo gerado por usuários em larga escala. Ebrahimi e Soltanifar (2024) exploraram o papel da plataforma na disseminação de notícias falsas, revelando os mecanismos pelos quais a desinformação se espalha em ecossistemas digitais. Por fim, Admassie e Melesse (2024) investigaram o potencial do Telegram em facilitar a pesquisa médica colaborativa, especialmente em regiões remotas, destacando sua adaptabilidade.

Seja na mobilização política, na educação ou na análise de comportamentos digitais, o Telegram serve como um facilitador-chave para o avanço acadêmico e o engajamento profissional. As capacidades técnicas da plataforma e as ferramentas analíticas desenvolvidas em torno dela continuarão expandindo os limites da pesquisa, oferecendo novos insights sobre fenômenos sociais e digitais complexos.



Com base nas aplicações destacadas na literatura acadêmica, este paper apresenta o impacto prático do código de extração do Telegram desenvolvido para fins de pesquisa. Ao possibilitar a coleta e a análise de dados de grupos e canais do Telegram, a ferramenta tem sido fundamental para o avanço de estudos empíricos em várias áreas. Notavelmente, ela apoiou investigações sobre dinâmicas de desinformação, discursos políticos e convergência temática em comunidades de teorias da conspiração. Abaixo, destacamos exemplos-chave em que essa ferramenta contribuiu para pesquisas inovadoras, demonstrando sua versatilidade e relevância ao abordar questões prementes da sociedade e da academia.

Para listar alguns exemplos de uso do código, ele foi utilizado no artigo "Informational Co-option against Democracy: Comparing Bolsonaro's Discourses about Voting Machines with the Public Debate" (Silva & Oliveira, 2023). Também foi empregado em "Institutional Denialism From the President's Speeches to the Formation of the Early Treatment Agenda (Off Label) in the COVID-19 Pandemic in Brazil" (Silva, 2023b). Além disso, o código facilitou a pesquisa "Catalytic Conspiracism: Exploring Persistent Homologies Time Series in the Dissemination of Disinformation in Conspiracy Theory Communities on Telegram" (Rocha, Silva & Mielli, 2024) e "Conspiratorial Convergence: Comparing Thematic Agendas Among Conspiracy Theory Communities on Telegram Using Topic Modeling" (Silva & Máximo, 2024). Também foi crucial em "Informational Disorder and Institutions Under Attack: How Did Former President Bolsonaro's Narratives Against the Brazilian Judiciary Between 2019 and 2022 Manifest?" (Silva, Castro, Santos, 2023).

Adicionalmente, o código foi empregado em várias notas técnicas, como em "Technical Note #16 – Disinformation about Electronic Voting Machines Persists Outside Election Periods" (Silva, 2023c) e "Technical Note #18 – Electoral Fraud Discourse in Argentina on Telegram and Twitter" (Silva & Ortellado, 2023). Também contribuiu para a análise da nota técnica "Bashing and Praising Public Servants and Bureaucrats During the Bolsonaro Government (2019 - 2022)" (Silva, Lotta, Fernandez, Seidi & Oliveira, 2024) e foi utilizado em "Technical Note 2: The Digital Territory of Milei's Followers: From Commerce to Politics" (Vargas, Hernández, Silva, Foltran, 2023).

Essa ferramenta também foi utilizada em uma série de sete artigos publicados no ArXiv sobre comunidades de teorias da conspiração, consolidando análises sobre o tema no Brasil: 1.) "Antivax and off-label medication communities on Brazilian Telegram: between esotericism as a gateway and the monetization of false miraculous cures" (Silva, 2024a); 2.)



"Climate change denial and anti-science communities on Brazilian Telegram: climate disinformation as a gateway to broader conspiracy networks" (Silva, 2024d); 3.) "Anti-woke agenda, gender issues, revisionism and hate speech communities on Brazilian Telegram: from harmful reactionary speech to the crime of glorifying Nazism and Hitler" (Silva, 2024c); 4.) "Apocalypse, survivalism, occultism and esotericism communities on Brazilian Telegram: when faith is used to sell quantum courses and open doors to harmful conspiracy theories" (Silva, 2024b); 5.) "UFO, universe, reptilians and creatures communities on Brazilian Telegram: when the sky is not the limit and conspiracy theories seek answers beyond humanity" (Silva, 2024h); 6.) "Flat-earth communities on Brazilian Telegram: when faith is used to question the existence of gravity as a physics phenomenon" (Silva, 2024e); 7.) "New world order, globalism and QAnon communities on Brazilian Telegram: how conspiracism opens doors to more harmful groups" (Silva, 2024g).

Além disso, a ferramenta foi aplicada em contextos de tomada de decisão procedimental, como durante a Comissão Parlamentar Mista de Inquérito sobre os eventos de 8 de janeiro de 2023. Os dados coletados por meio dessa ferramenta foram fundamentais para embasar os mandados de intimação de indivíduos envolvidos na invasão criminosa a Brasília, Brasil, nessa data (Senado Federal, 2023; Psol na Câmara, 2023). A capacidade de coletar e analisar dados em grande escala sistematicamente permitiu às autoridades identificar e direcionar atores específicos no âmbito investigativo, demonstrando a utilidade da ferramenta em processos legais e políticos. Adicionalmente, essa tecnologia tem sido utilizada para rastrear criminosos e abordar crimes cibernéticos na Itália. Pesquisadores afiliados ao DarkLab of Ethical Hacking destacaram seu papel na identificação e acusação de ofensores envolvidos em atividades online maliciosas, particularmente por meio da análise de vazamentos de dados e ameaças cibernéticas (RedHot Cyber, 2024). Esses exemplos reforçam a versatilidade da ferramenta, apontando a sua eficácia em aplicações diversas, que vão desde a segurança pública até investigações de crimes cibernéticos, enquanto avançam nas áreas de forense digital e justiça procedimental.

## 2. Uso passo a passo

Apenas algumas dicas introdutórias:



- ➔ **Limite de tempo no Google Colab:** O runtime do Google Colab normalmente encerra após cerca de 6 horas e 20 minutos (22.800 segundos). Assim, é recomendável definir um limite de tempo dentro desse intervalo para evitar interrupções.
- ➔ **Soft ban da API do Telegram:** A API do Telegram geralmente aplica um soft ban de 24 horas após a extração de dados de mais de 200 canais ou grupos. No entanto, não parece haver limite para a quantidade de mensagens extraídas de um número menor de comunidades. Para evitar o soft ban, extraia grandes volumes de conteúdo em blocos de até 150-200 comunidades por vez, independentemente de serem dados referentes a meses inteiros ou apenas alguns dias.
- ➔ **Uso do Google Colab para operações assíncronas:** Uma vantagem do Google Colab é a capacidade de executar operações async sem a necessidade de defini-las dentro de um async def. Caso planeje usar o PyCharm ou outro IDE, considere adaptar o código com uma função async def.
- ➔ **Manipulação de JSON na coluna 'Comments List':** A coluna 'Comments List' armazena comentários em formato de lista JSON. Certifique-se de decodificar esse JSON ao converter os dados para uma planilha ou apresentá-los. É importante lembrar que, caso a lista de comentários seja muito longa, ela não caberá em uma célula de Excel. Por esse motivo (entre outras vantagens), o arquivo em formato Parquet com JSON é fornecido, para evitar quebras de conteúdo.

### *I. [ Obrigatório ] Configure suas credenciais uma única vez*

O script foi projetado para facilitar a extração de dados de canais ou grupos do Telegram por meio de um processo estruturado. No primeiro passo, os usuários precisam configurar suas credenciais, o que é uma configuração realizada apenas uma vez. Isso envolve o preenchimento das seguintes informações obrigatórias: username, phone, api_id e api_hash. O username deve ser fornecido sem o símbolo "@", e o número de telefone deve estar no formato internacional, por exemplo, "+5511999999999". Os campos api_id e api_hash são essenciais para autenticação e só podem ser gerados por meio da página de criação de aplicativos do Telegram, acessível em https://my.telegram.org/apps. Após preencher esses campos, basta executar a célula para inicializar as credenciais. Essas configurações não precisam ser atualizadas, a menos que o usuário decida alterar sua conta do Telegram.



**Figura 1 -** Primeira célula a ser avaliada

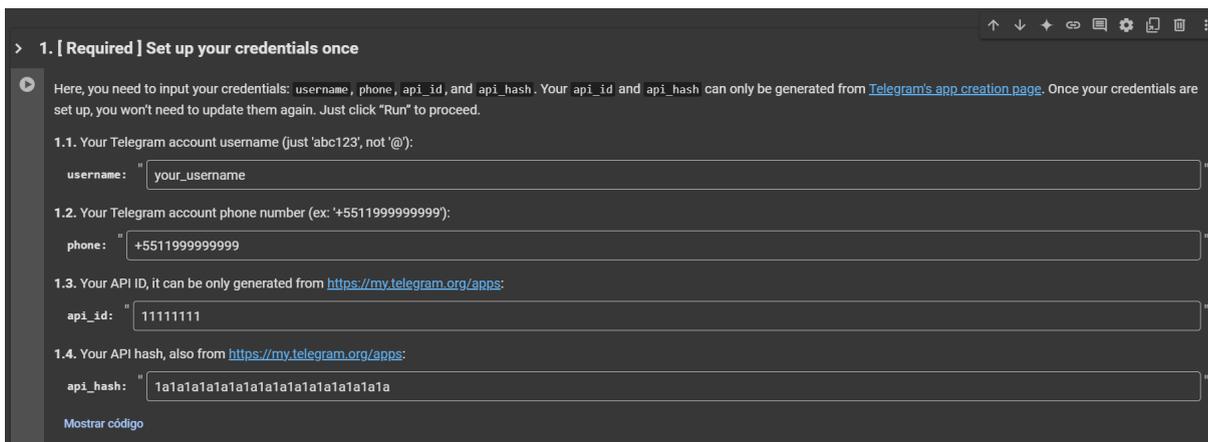

**Fonte: Elaboração própria (2024).**

## II. [ Obrigatório ] Ajuste sempre que quiser utilizá-lo

No segundo passo, os usuários configuram os parâmetros para cada sessão de extração. Esta seção permite definir as condições específicas para a coleta de dados. O primeiro parâmetro é a lista de canais ou grupos do Telegram a serem extraídos. Esses devem ser escritos como @ChannelName ou o URL completo, como https://t.me/ChannelName, separados por vírgulas. Links da versão web do Telegram, como "https://web.telegram.org", não devem ser usados. O segundo parâmetro é o intervalo de datas, definido por date_min e date_max, que especifica o período do qual as mensagens serão coletadas. As datas devem estar no formato AAAA-MM-DD. Os usuários também precisam nomear o arquivo de saída para os dados coletados, garantindo clareza para uso futuro. Um campo opcional de busca por palavras-chave permite filtrar mensagens por termos específicos; se deixado em branco, todas as mensagens dentro do intervalo especificado serão extraídas. Além disso, os usuários podem definir um número máximo de mensagens a serem extraídas, garantindo que o processo permaneça gerenciável. A duração do timeout, que é o tempo máximo que o script será executado, também deve ser definida. Este valor não deve exceder seis horas (21.600 segundos), pois sessões do Google Colab são automaticamente encerradas após esse período. Por fim, os usuários escolhem o formato do arquivo de saída: excel para maior acessibilidade ou parquet para processamento avançado de dados.



**Figura 2 -** Segunda célula com parâmetros de busca

**Fonte: Elaboração própria (2024).**

## III. [ Obrigatório ] Inicie a extração de dados do Telegram

O terceiro passo consiste em executar a extração de dados. Durante esta etapa, o script inicia a coleta de dados com base nos parâmetros definidos anteriormente. Os usuários devem estar atentos ao fato de que o Telegram pode solicitar um código de verificação durante o processo, que precisará ser inserido prontamente. O script coleta mensagens juntamente com seus metadados, como visualizações, compartilhamentos e reações, além de comentários, quando disponíveis. Atualizações de progresso são exibidas ao longo da sessão, mostrando a porcentagem concluída, o tempo decorrido e o tempo estimado restante. Para minimizar o risco de perda de dados em sessões prolongadas, arquivos de backup são criados automaticamente a cada 1.000 mensagens processadas. Esses arquivos de backup são salvos no formato de saída especificado, garantindo que os usuários tenham acesso aos resultados intermediários mesmo que a sessão seja interrompida. Ao concluir o processo de extração, o script compila um arquivo final contendo todos os dados coletados, pronto para download.



**Figura 3 -** Terceira célula onde os Metadados são extraídos

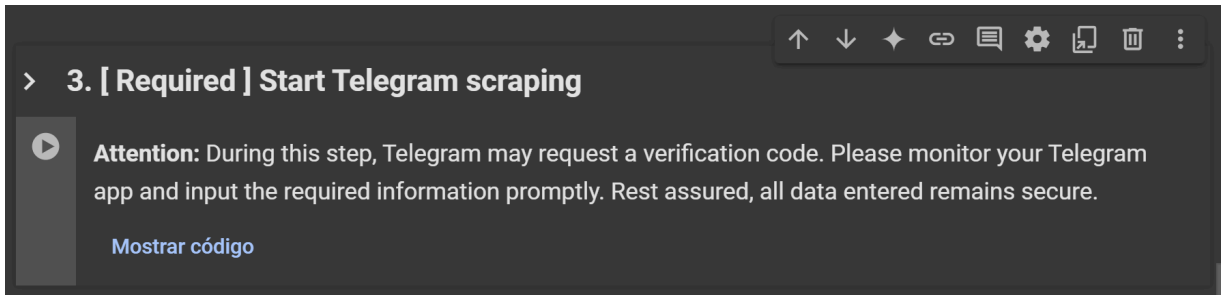

**Fonte: Elaboração própria (2024).**

## 3. Código

### *I. Configure suas credenciais uma única vez*

**Tabela 01.** Abordagens da primeira célula e código

| Abordagem | Código |
|---|---|
| Ele utiliza a biblioteca Telethon, uma ferramenta robusta para interagir programaticamente com o Telegram. Os usuários são solicitados a inserir seu username, número de telefone, api_id e api_hash, credenciais essenciais geradas na página de criação de aplicativos do Telegram (https://my.telegram.org/apps). Esses detalhes estabelecem a conexão com a conta do usuário no Telegram. Além do Telethon, o script importa bibliotecas fundamentais como datetime para gerenciar carimbos de data e hora, pandas para manipulação de dados, json e re para processamento e limpeza de dados textuais, e time para rastrear a duração das execuções. O módulo files do Google Colab também é incluído para facilitar o download direto de arquivos para a máquina do usuário. Essa célula deve ser executada apenas uma vez, a menos que as credenciais precisem ser atualizadas. | ```python
# Install the Telethon library for Telegram API interactions
!pip install -q telethon

# Initial imports
from datetime import datetime, timezone
import pandas as pd
import time
import json
import re

# Telegram imports
from telethon.sync import TelegramClient

# Google Colab imports
from google.colab import files

# Setup / change only the first time you use it
# @markdown **1.1.** Your Telegram account username (just 'abc123', not '@'):
username = 'your_username' # @param {type:"string"}
# @markdown **1.2.** Your Telegram account phone number (ex: '+5511999999999'):
phone = '+5511999999999' # @param {type:"string"}
# @markdown **1.3.** Your API ID, it can be only generated from https://my.telegram.org/apps:
api_id = '11111111' # @param {type:"string"}
# @markdown **1.4.** Your API hash, also from https://my.telegram.org/apps:
api_hash = '1a1a1a1a1a1a1a1a1a1a1a1a1a1a1a1a' # @param {type:"string"}
``` |

**Fonte: Elaboração própria (2024).**

## II. Ajuste sempre que quiser utilizá-lo

**Quadro 02.** Abordagens da segunda célula e código

| Abordagem | Código |
|---|---|
| Oferece uma interface flexível para definir os parâmetros de cada sessão de extração de dados do Telegram, permitindo personalização total para atender a diversas necessidades de pesquisa. Essa célula utiliza bibliotecas essenciais como datetime para gerenciar intervalos de datas e pandas para a manipulação estruturada de dados, garantindo o processamento preciso das configurações definidas pelo usuário. Os usuários são orientados a inserir uma lista de canais ou grupos do Telegram no formato @ChannelName ou URLs completas, como https://t.me/ChannelName. Técnicas dinâmicas de manipulação de strings (.split, .strip) tratam da separação e limpeza dos valores inseridos, enquanto links de plataformas incompatíveis, como https://web.telegram.org, são excluídos para assegurar uma interação fluida com a API do Telegram. Essa célula permite especificar um intervalo de datas (date_min e date_max) para filtrar mensagens de períodos específicos. As datas são automaticamente convertidas para UTC usando datetime.fromisoformat, tornando esse recurso particularmente valioso para análises sensíveis ao tempo, como o exame de discussões relacionadas a eventos políticos ou políticas públicas. Os usuários também atribuem um nome ao arquivo de saída (file_name) para organizar de forma eficaz múltiplos conjuntos de dados. Um filtro opcional de palavras-chave pode ser aplicado para focar em mensagens que contenham termos específicos, enquanto deixar o campo em branco permite a extração de todas as mensagens disponíveis. Para garantir que o processo permaneça gerenciável, os usuários podem definir um número máximo de mensagens a serem extraídas (max_t_index) e estabelecer um limite de tempo em segundos | ```python
# Setup / change every time to define scraping parameters

# @markdown **2.1.** Here you put the name of the channel or group that you want to scrape, as an example, play: '@LulanoTelegram' or 'https://t.me/LulanoTelegram'. Do not use: 'https://web.telegram.org/a/#-1001249230829' or '-1001249230829'. **Just write the `channel names` always separated by commas (,):**
channels = "@LulanoTelegram, @jairbolsonarobrasil, @Other_Channel_Name" # @param {type:"string"}
channels = [channel.strip() for channel in channels.split(",")]

# @markdown **2.2.** Here you can select the `time window` you would like to extract data from the listed communities:
date_min = '2024-10-15' # @param {type:"date"}
date_max = '2025-01-15' # @param {type:"date"}

date_min = datetime.fromisoformat(date_min).replace(tzinfo=timezone.utc)
date_max = datetime.fromisoformat(date_max).replace(tzinfo=timezone.utc)

# @markdown **2.3.** Choose a `name` for the final file you want to download as output:
file_name = 'Test' # @param {type:"string"}

# @markdown **2.4.** `Keyword` to search, **leave empty if you want to extract all messages from the channel(s):**
key_search = '' # @param {type:"string"}

# @markdown **2.5.** **Maximum** `number of messages` to scrape (only use if you want a specific limit, otherwise leave a high number to scrape everything):
max_t_index = 1000000   # @param {type:"integer"}

# @markdown **2.6.** `Timeout in seconds` (never leave it longer than 6 hours, that is 21600 seconds, as Google Colab deactivates itself after that time):
time_limit = 21600 # @param {type:"integer"}

# @markdown **2.7.** Choose the format of the final file you want to download. If you are a first-time user, choose `Excel`. If you have advanced skills, you can use `Parquet`:
File = 'excel' # @param ["excel", "parquet"]
``` |



| Abordagem | Código |
|---|---|
| (timeout). Esse timeout assegura que a sessão seja concluída dentro da restrição de execução de seis horas do Google Colab (21.600 segundos). A célula também oferece a opção de escolher entre dois formatos de arquivo de saída: Excel, para maior acessibilidade e facilidade de uso, ou Parquet, que é otimizado para lidar com grandes volumes de dados e amplamente utilizado em fluxos de trabalho de big data. | |

**Fonte:** Elaboração própria (2024).

### III. Inicie a extração de dados do Telegram

**Quadro 03.** Abordagens da terceira célula e código

| Abordagem | Código |
|---|---|
| Executa o processo principal de coleta de dados integrando as configurações e bibliotecas definidas nas etapas anteriores. Ele utiliza o TelegramClient da biblioteca Telethon para interagir programaticamente com a API do Telegram.<br><br>O processo de extração é assíncrono, garantindo desempenho eficiente mesmo ao lidar com grandes volumes de dados em plataformas com recursos limitados, como o Google Colab. A célula inicializa variáveis globais, como uma lista para armazenar os dados coletados (data), um contador de mensagens (t_index) e um carimbo de data e hora para acompanhar o início da sessão (start_time). Esses elementos fundamentais facilitam o acompanhamento do progresso e garantem o gerenciamento organizado dos dados ao longo da sessão.<br><br>Uma das principais funcionalidades dessa célula é sua capacidade de processar mensagens e seus metadados, incluindo visualizações, reações, compartilhamentos e quaisquer mídias associadas. | ```python
data = []  # List to store scraped data
t_index = 0  # Tracker for the number of messages processed
start_time = time.time()  # Record the start time for the scraping session

# Function to remove invalid XML characters from text
def remove_unsupported_characters(text):
    valid_xml_chars = (
        "[^\u0009\u000A\u000D\u0020-\uD7FF\uE000-\uFFFD"
        "\U00010000-\U0010FFFF]"
    )
    cleaned_text = re.sub(valid_xml_chars, '', text)
    return cleaned_text

# Function to format time in days, hours, minutes, and seconds
def format_time(seconds):
    days = seconds // 86400
    hours = (seconds % 86400) // 3600
    minutes = (seconds % 3600) // 60
    seconds = seconds % 60
    return f'{int(days):02}:{int(hours):02}:{int(minutes):02}:{int(seconds):02}'

# Function to print progress of the scraping process
def print_progress(t_index, message_id, start_time, max_t_index):
    elapsed_time = time.time() - start_time
    current_progress = t_index / (t_index + message_id) if (t_index + message_id) <= max_t_index else t_index / max_t_index
``` |

| Abordagem | Código |
|---|---|
| Utilizando um loop assíncrono, o script percorre mensagens dentro do intervalo de datas especificado, também capturando e processando comentários vinculados a mensagens individuais. Ele emprega funções personalizadas, como remove_unsupported_characters, para limpar textos, removendo caracteres XML inválidos e garantindo que o conjunto final de dados esteja livre de erros de codificação. Para melhorar a experiência do usuário, o progresso é exibido em tempo real por meio de funções como format_time e print_progress, que fornecem atualizações sobre o tempo decorrido, a porcentagem de conclusão e o tempo estimado restante. A célula incorpora medidas robustas de segurança de dados, gerando automaticamente arquivos de backup a cada 1.000 mensagens processadas. Esses backups recebem nomes dinâmicos, que incluem o número de mensagens coletadas e o identificador do canal, garantindo rastreabilidade. Dependendo da preferência do usuário, os arquivos são salvos em formato Excel ou Parquet, oferecendo flexibilidade para usuários casuais e avançados. Caso erros ocorram durante o processamento de mensagens ou comentários, as exceções são tratadas de forma controlada, permitindo que o script continue sua execução sem interrupções significativas. O gerenciamento do tempo é outro recurso crítico dessa célula. O script acompanha a duração da sessão, encerrando automaticamente caso o limite de timeout predefinido seja alcançado, respeitando as restrições de execução do Google Colab. Ao final da sessão de extração, o script compila todos os dados coletados em um arquivo final e o disponibiliza para download direto. Isso é facilitado pelo módulo files do Google Colab, simplificando a transferência dos resultados para o dispositivo local do usuário. A arquitetura da célula demonstra uma integração sofisticada de programação assíncrona, acompanhamento de progresso em | ```python
    percentage = current_progress * 100
    estimated_total_time = elapsed_time / current_progress
    remaining_time = estimated_total_time - elapsed_time

    elapsed_time_str = format_time(elapsed_time)
    remaining_time_str = format_time(remaining_time)

    print(f'Progress: {percentage:.2f}% | Elapsed Time: {elapsed_time_str} | Remaining Time: {remaining_time_str}')

# Normalize File variable to avoid issues
File = re.sub(r'[^a-z]', '', File.lower())  # Converts to lowercase and removes non-alphabetic characters

# Scraping process
for channel in channels:
    if t_index >= max_t_index:
        break

    if time.time() - start_time > time_limit:
        break

    loop_start_time = time.time()

    try:
        c_index = 0
        async with TelegramClient(username, api_id, api_hash) as client:
            async for message in client.iter_messages(channel, search=key_search):
                try:
                    if date_min <= message.date <= date_max:

                        # Process comments of the message
                        comments_list = []
                        try:
                            async for comment_message in client.iter_messages(channel, reply_to=message.id):
                                comment_text = comment_message.text.replace("'", '"')

                                comment_media = 'True' if comment_message.media else 'False'

                                comment_emoji_string = ''
                                if comment_message.reactions:
                                    for reaction_count in comment_message.reactions.results:
                                        emoji = reaction_count.reaction.emoticon
                                        count = str(reaction_count.count)
``` |



| Abordagem | Código |
|---|---|
| tempo real e gerenciamento de arquivos, garantindo operação eficiente mesmo em cenários complexos. Seu design modular suporta fácil adaptação para estudos futuros, como análises de redes sociais, acompanhamento de desinformação ou investigação de dinâmicas comunitárias. | ```python
comment_emoji_string += emoji + " " + count + " "
                                    comment_date_time = comment_message.date.strftime('%Y-%m-%d %H:%M:%S')

comments_list.append({
                                        'Type': 'comment',
                                        'Comment Group': channel,
                                        'Comment Author ID': comment_message.sender_id,
                                        'Comment Content': comment_text,
                                        'Comment Date': comment_date_time,
                                        'Comment Message ID': comment_message.id,
                                        'Comment Author': comment_message.post_author,
                                        'Comment Views': comment_message.views,
                                        'Comment Reactions': comment_emoji_string,
                                        'Comment Shares': comment_message.forwards,
                                        'Comment Media': comment_media,
                                        'Comment Url': f'https://t.me/{channel}/{message.id}?comment={comment_message.id}'.replace('@', ''),
                                    })
                            except Exception as e:
                                comments_list = []
                                print(f'Error processing comments: {e}')

                            # Process the main message
                            media = 'True' if message.media else 'False'

                            emoji_string = ''
                            if message.reactions:
                                for reaction_count in message.reactions.results:
                                    emoji = reaction_count.reaction.emoticon
                                    count = str(reaction_count.count)
                                    emoji_string += emoji + " " + count + " "

                            date_time = message.date.strftime('%Y-%m-%d %H:%M:%S')
                            cleaned_content = remove_unsupported_characters(message.text)
                            cleaned_comments_list = remove_unsupported_characters(json.dumps(comments_list))

                            data.append({
                                'Type': 'text',
                                'Group': channel,
``` |

| Abordagem | Código |
|---|---|
|  | ```python
                                'Author ID': message.sender_id,
                                'Content': cleaned_content,
                                'Date': date_time,
                                'Message ID': message.id,
                                'Author': message.post_author,
                                'Views': message.views,
                                'Reactions': emoji_string,
                                'Shares': message.forwards,
                                'Media': media,
                                'Url': f'https://t.me/{channel}/{message.id}'.replace('@', ''),
                                'Comments List': cleaned_comments_list,
                            })

                            c_index += 1
                            t_index += 1

                            # Print progress
                            print(f'{"-" * 80}')
                            print_progress(t_index, message.id, start_time, max_t_index)
                            current_max_id = min(c_index + message.id, max_t_index)
                            print(f'From {channel}: {c_index:05} contents of {current_max_id:05}')
                            print(f'Id: {message.id:05} / Date: {date_time}')
                            print(f'Total: {t_index:05} contents until now')
                            print(f'{"-" * 80}\n\n')

                            if t_index % 1000 == 0:
                                if File == 'parquet':
                                    backup_filename = f'backup_{file_name}_until_{t_index:05}_{channel}_ID{message.id:07}.parquet'
                                    pd.DataFrame(data).to_parquet(backup_filename, index=False)
                                elif File == 'excel':
                                    backup_filename = f'backup_{file_name}_until_{t_index:05}_{channel}_ID{message.id:07}.xlsx'
                                    pd.DataFrame(data).to_excel(backup_filename, index=False, engine='openpyxl')

                            if t_index >= max_t_index:
                                break

                            if time.time() - start_time > time_limit:
                                break

                        elif message.date < date_min:
                            break
``` |

| Abordagem | Código |
|---|---|
| | ```python
                except Exception as e:
                    print(f'Error processing message: {e}')

            print(f'\n\n##### {channel} was ok with {c_index:05} posts #####\n\n')

            df = pd.DataFrame(data)
            if File == 'parquet':
                partial_filename = f'complete_{channel}_in_{file_name}_until_{t_index:05}.parquet'
                df.to_parquet(partial_filename, index=False)
            elif File == 'excel':
                partial_filename = f'complete_{channel}_in_{file_name}_until_{t_index:05}.xlsx'
                df.to_excel(partial_filename, index=False, engine='openpyxl')
            # files.download(partial_filename)

        except Exception as e:
            print(f'{channel} error: {e}')

        loop_end_time = time.time()
        loop_duration = loop_end_time - loop_start_time

        if loop_duration < 60:
            time.sleep(60 - loop_duration)

print(f'\n{"-" * 50}\n#Concluded! #{t_index:05} posts were scraped!\n{"-" * 50}\n\n\n\n')
df = pd.DataFrame(data)
if File == 'parquet':
    final_filename = f'FINAL_{file_name}_with_{t_index:05}.parquet'
    df.to_parquet(final_filename, index=False)
elif File == 'excel':
    final_filename = f'FINAL_{file_name}_with_{t_index:05}.xlsx'
    df.to_excel(final_filename, index=False, engine='openpyxl')
files.download(final_filename)
``` |

**Fonte: Elaboração própria (2024).**

## 4. Conclusões

A ferramenta **TelegramScrap** propõe uma solução para a extração, organização e análise de dados de canais e grupos do Telegram, atendendo à demanda por ferramentas para navegar por plataformas digitais. Ao integrar funcionalidades intuitivas, a ferramenta conecta a disponibilidade de dados brutos a insights acionáveis. Ela capacita pesquisadores, jornalistas e analistas a explorar comportamentos digitais complexos, acompanhar campanhas de desinformação e investigar padrões temáticos em comunidades online.

Um dos principais pontos fortes da ferramenta é sua adaptabilidade, atendendo a diversas necessidades e metodologias de pesquisa. A arquitetura modular permite personalização fluida, possibilitando que os usuários ajustem parâmetros de extração, formatos de saída e focos analíticos de acordo com seus objetivos específicos. O acompanhamento de progresso em tempo real e os backups automáticos aumentam a confiabilidade, tornando-a um recurso valioso para estudos longitudinais e extração de grandes volumes de dados. Além disso, a ferramenta incorpora uma abordagem ética à coleta de dados, enfatizando transparência, responsabilidade do usuário e conformidade com os regulamentos da plataforma. Ela promove o uso responsável ao destacar limitações como restrições da API, considerações de tempo de execução e a importância de respeitar a privacidade dos usuários. Esses mecanismos garantem que a ferramenta seja não apenas eficaz, mas também alinhada a padrões éticos e à conformidade legal.

Seu uso é altamente incentivado e recomendado para pesquisas acadêmicas e científicas, análises de conteúdo, análises de sentimento e análises de discurso. Embora seja gratuita para uso e modificação, a responsabilidade por seu uso e quaisquer alterações recai sobre o usuário. Sinta-se à vontade para explorar, utilizar e adaptar o código às suas necessidades, mas certifique-se de cumprir os termos de serviço do Telegram e as regulamentações de privacidade de dados. Essa ferramenta é lançada sob uma licença gratuita e de código aberto. Ao utilizá-la ou modificá-la, é importante fornecer o devido crédito e citação. Referenciar a ferramenta em sua pesquisa é apreciado e contribui para seu desenvolvimento e aprimoramento contínuos.

## 5. Referências

### *5.1. Ferramenta*

### *5.2. Bibliografia*

### 5.3. *Alguns dos papers que utilizaram o código*

## 6. Biografia do autor

**Ergon Cugler de Moraes Silva** possui Mestrado em Administração Pública e Governo (FGV), MBA em Ciência de Dados e Analytics (USP), Bacharelado em Gestão de Políticas Públicas (USP) e atualmente cursa Pós-Graduação em Data Science for Social and Business Analytics na Universitat de Barcelona. É associado à More in Common (Brasil) e colabora com o Observatório Interdisciplinar de Políticas Públicas (OIPP USP), o Grupo de Estudos sobre Tecnologia e Inovações na Gestão Pública (GETIP USP), o Monitor do Debate Político no Meio Digital (Monitor USP) e o Grupo de Trabalho em Estratégia, Dados e Soberania do Grupo de Estudos e Pesquisas em Segurança Internacional do Instituto de Relações Internacionais da Universidade de Brasília (GEPSI UnB). É também pesquisador no Instituto Brasileiro de Informação em Ciência e Tecnologia (IBICT), onde atua pelo Governo

Federal em estratégias de enfrentamento à desinformação. São Paulo, São Paulo, Brasil. Website: https://ergoncugler.com/.